\documentclass[aps,prl,preprint,superscriptaddress,nopacs]{revtex4-1}
\usepackage{graphicx}
\usepackage[colorlinks=true,citecolor=blue,linkcolor=magenta]{hyperref}
\usepackage{amsmath}
\usepackage{siunitx}
\hypersetup{
pdfauthor = {Patrick Maletinsky},
colorlinks = true, linkcolor = blue, urlcolor=blue, bookmarksnumbered =  true}

\newcommand{\CrO}{Cr$_2$O$_3$}
\newcommand{\Tsub}[1]{T_{\mathrm{#1}}}

\begin{document}

%TC:ignore %Marker for wordcount

%Title of paper
\title{Nanomagnetism of magnetoelectric granular thin-film antiferromagnets}
\author{Patrick Appel}
\thanks{These authors contributed equally}
\affiliation{Department of Physics, University of Basel, Klingelbergstrasse 82, Basel CH-4056, Switzerland}
\author{Brendan J. Shields}
\thanks{These authors contributed equally}
\affiliation{Department of Physics, University of Basel, Klingelbergstrasse 82, Basel CH-4056, Switzerland}
\author{Tobias Kosub}
\affiliation{Helmholtz-Zentrum Dresden-Rossendorf e.V., Institute of Ion Beam Physics and Materials Research, 01328 Dresden, Germany}
\affiliation{Institute for Integrative Nanosciences, Institute for Solid State and Materials Research (IFW Dresden e.V.), 01069 Dresden, Germany.}
\author{Ren{\'e} H{\"u}bner}
\affiliation{Helmholtz-Zentrum Dresden-Rossendorf e.V., Institute of Ion Beam Physics and Materials Research, 01328 Dresden, Germany}
\author{J{\"u}rgen Fa{\ss}bender}
\affiliation{Helmholtz-Zentrum Dresden-Rossendorf e.V., Institute of Ion Beam Physics and Materials Research, 01328 Dresden, Germany}
\author{Denys Makarov}
\affiliation{Helmholtz-Zentrum Dresden-Rossendorf e.V., Institute of Ion Beam Physics and Materials Research, 01328 Dresden, Germany}
\affiliation{Institute for Integrative Nanosciences, Institute for Solid State and Materials Research (IFW Dresden e.V.), 01069 Dresden, Germany.}
\author{Patrick Maletinsky}
\email[]{patrick.maletinsky@unibas.ch}
\affiliation{Department of Physics, University of Basel, Klingelbergstrasse 82, Basel CH-4056, Switzerland}

%\homepage[]{Your web page}
%\thanks{}

\date{\today}

\begin{abstract} 
Antiferromagnets have recently emerged as attractive platforms for spintronics applications, offering fundamentally new functionalities compared to their ferromagnetic counterparts. 
While nanoscale thin film materials are key to the development of future antiferromagnetic spintronics technologies, experimental tools to explore such films on the nanoscale are still sparse.  
Here, we offer a solution to this technological bottleneck, by addressing the ubiquitous surface magnetisation of magnetoelectic antiferromagnets in a granular thin film sample on the nanoscale using single-spin magnetometry in combination with spin-sensitive transport experiments. 
Specifically, we quantitatively image the evolution of individual nanoscale antiferromagnetic domains in $200$-nm thin-films of \CrO{} in real space and across the paramagnet-to-antiferromagnet phase transition.  
These experiments allow us to discern key properties of the \CrO{} thin film, including the mechanism of domain formation and the strength of exchange coupling between individual grains comprising the film.  
Our work offers novel insights into \CrO{}'s magnetic ordering mechanism and establishes single spin magnetometry as a novel, widely applicable tool for nanoscale addressing of antiferromagnetic thin films. 
\end{abstract}

\maketitle

The combination of long-range magnetic order, vanishing macroscopic magnetisation, and strong exchange coupling between spin sublattices renders antiferromagnetic (AF) systems distinctly different from their ferromagnetic counterparts. 
Striking examples include terahertz spin dynamics, negligible domain cross-talk, and the ability to electrically control magnetic order of magnetoelectric AFs\,\cite{Heron2014,Kosub2017}, all of which offer highly attractive prospects for AF-based spintronics\,\cite{Jungwirth2016,Baltz2016,Seki2015} and data storage\,\cite{Wadley2016,Kosub2017,Loth2012}.
While the groundwork for the understanding of AFs was established through studies on bulk single crystals, thin-film AF materials, which typically show nanoscale granularity,  are required to translate the attractive physical properties of AFs into viable technologies.
As evidenced by the decades-long development of ferromagnetic memory media\,\cite{Piramanayagam2007}, the properties of such films can deviate significantly from their bulk parent. 
In particular, inter-granular exchange coupling is a decisive parameter, determining quantities such as size and switching speeds of memory bits.
The development of AF thin films into future spintronics technologies therefore hinges on the control and optimisation of such quantities and requires methods for quantitative, nanoscale studies of AF thin film materials\,\cite{Schmid2010,Wu2011,OGrady2010}. 

%TC:ignore 
\begin{figure}
\includegraphics[width =8.6cm]{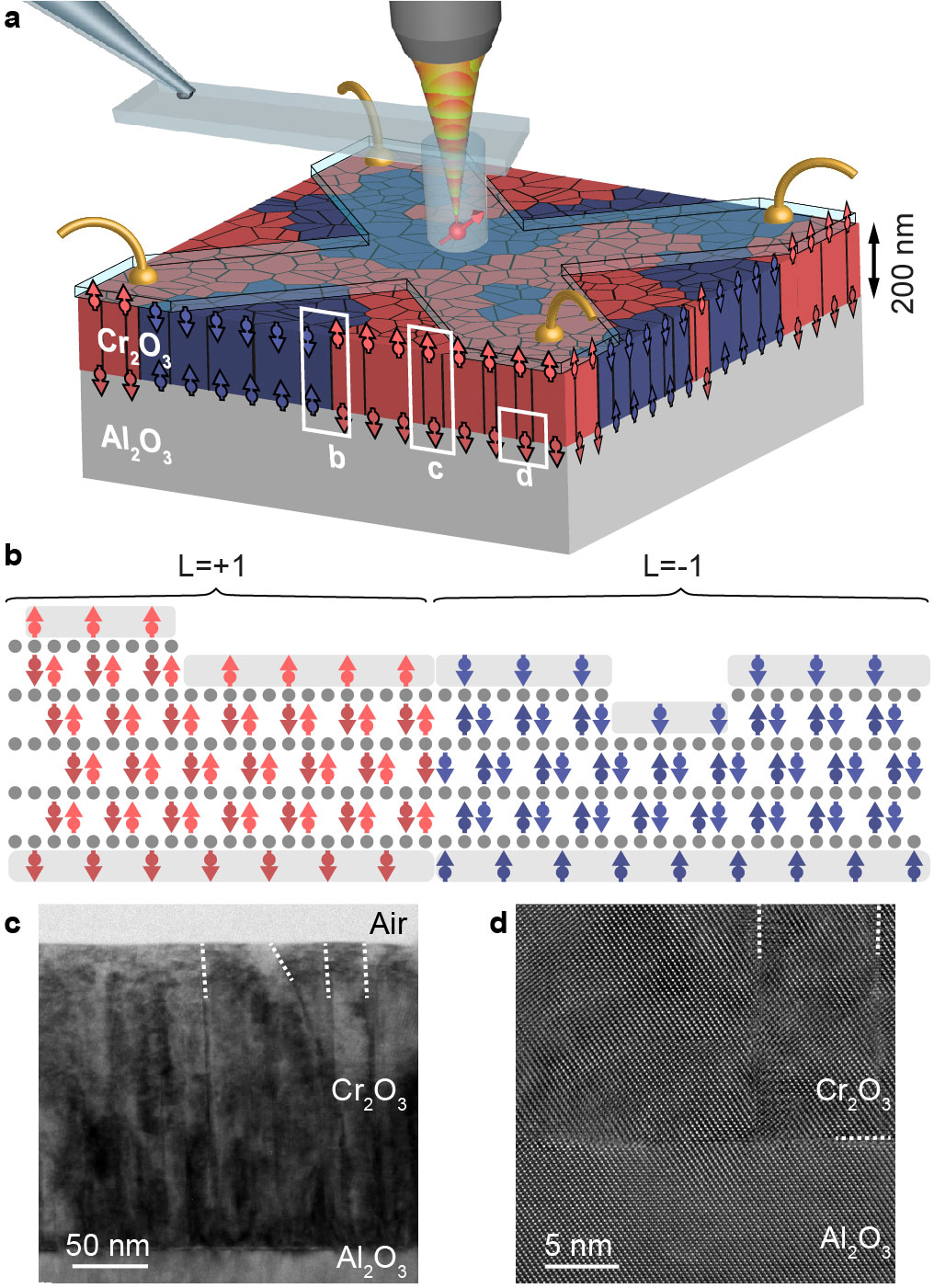}
\caption{{\bf Schematic of experiment and thin film antiferromagnetic sample.} {\bf a} A thin film sample of antiferromagnetic \CrO{} is examined using a combination of scanning single spin magnetometry (red arrow) and zero-offset Hall magnetometry (light blue cross with golden leads). 
{\bf b} \CrO{} is a bulk antiferromagnet with a roughness-insensitive, non-zero surface magnetisation linked to the underlying order parameter.
{\bf c} and  {\bf d} Cross-sectional transmission electron microscopy images of the sample. The images show the typical, columnar grains comprising the \CrO{} film (representative grain-boundaries highlighted by white dashed lines), and the high crystalline quality of the film. 
\label{Fig:Fig1}}
\end{figure}
%TC:endignore 

Here, we introduce a novel approach for characterising nanoscale magnetic properties of thin film AFs (Fig.\,\ref{Fig:Fig1}a) using Nitrogen Vacancy (NV) single spin magnetometry\,\cite{Rondin2014} -- an emerging, scanning probe-based quantum sensing technique operating under ambient conditions.
Our approach is based on the common feature of magnetoelectric AFs to exhibit non-zero boundary magnetisations\,\cite{Belashchenko2010,Andreev1996}, as a result of broken crystal symmetries on their surface. 
Therefore, measuring the weak magnetic stray fields emerging from this magnetisation enables access to the underlying AF order.
We apply this concept to the exemplary AF \CrO{}, in which we image and study nanoscale domain patterns across the paramagnet-AF phase transition with unprecedented signal-to-noise ratio and sub-$100~$nm spatial resolution. 
In combination with zero-offset anomalous Hall effect magnetometry\,\cite{Kosub2015}, our nanoscale magnetic sensing experiments reveal the formation of AF domains, whose existence we attribute to inhomogeneities and significant inter-granular exchange coupling in the thin film. 

Chromium oxide (\CrO{}) is a magnetoelectric AF and, due to its room temperature ordering and magnetoelectric switching capabilities\,\cite{Kosub2017,He2010,Ashida2014}, a key contender for future, AF-based magnetoelectric memory devices. 
Due to breaking of the crystal bulk symmetry, the (0001) surface of \CrO{} consists of a layer of Cr atoms that all belong to the same AF sublattice, and therefore present a roughness-insensitive, non-zero surface magnetisation, whose orientation is rigidly linked to the AF order parameter\,\cite{Belashchenko2010,He2010} (Fig.\,\ref{Fig:Fig1}b). 
This surface magnetisation, on the order of few $\mu_{\rm{B}}/$nm$^2$\,\cite{Andreev1996} (where $\mu_{\rm{B}}$ is the Bohr magneton) 
creates stray magnetic fields which, in the far-field, are quadrupolar in nature and decay with distance from the sample over lengthscales given by typical domain sizes or sample thickness\,\cite{Andreev1996,Astrov1996}.
Exploiting such stray fields to address thin film samples thus requires a sensitive magnetometer that can be brought in close proximity to the sample.

This non-trivial requirement is met by nanoscale, scanning NV magnetometry (Fig.\,\ref{Fig:Fig1}a)\,\cite{Maletinsky2012,Appel2016,Rondin2014}, a technique that has previously been used to study magnetism in thin films possessing an overall magnetic moment\,\cite{Tetienne2014,Gross2017}.
The technique operates by scanning a single electronic spin in close vicinity to a surface to measure the magnetic field $B_{\rm NV}$ along the fixed NV quantisation axis through the Zeeman splitting of the NV's optically detected electron spin resonance (see Methods). 
It operates under ambient conditions, offers quantitative sensing, nanoscale imaging and sensitivities sufficient to address individual electronic spins\,\cite{Grinolds2013}. These unique characteristics are ideally suited to address the nanoscale properties of AF systems targeted here.

The samples we investigate consist of $200$-nm-thick \CrO{} films grown on a $c$-cut sapphire substrate  
(see Methods). 
The obtained film is granular with $\approx50~$nm-sized, columnar grains with smooth surfaces (Fig.\,\ref{Fig:Fig1}c) and very high crystallinity within each grain (Fig.\,\ref{Fig:Fig1}d).  
For NV magnetometry, the sample is mounted on a Peltier element to control  temperatures to within $\pm0.1\,$K during measurements. We prepare the film by heating it well above the AF-to-paramagnet transition ($T_{\rm Neel}=308~$K in bulk\,\cite{Wu2011}), and then cooling back into the AF state 
in zero external magnetic field 
to induce a spontaneously formed pattern of AF domains\,\cite{Kosub2017}.

\begin{figure}
\includegraphics[width =8.6cm]{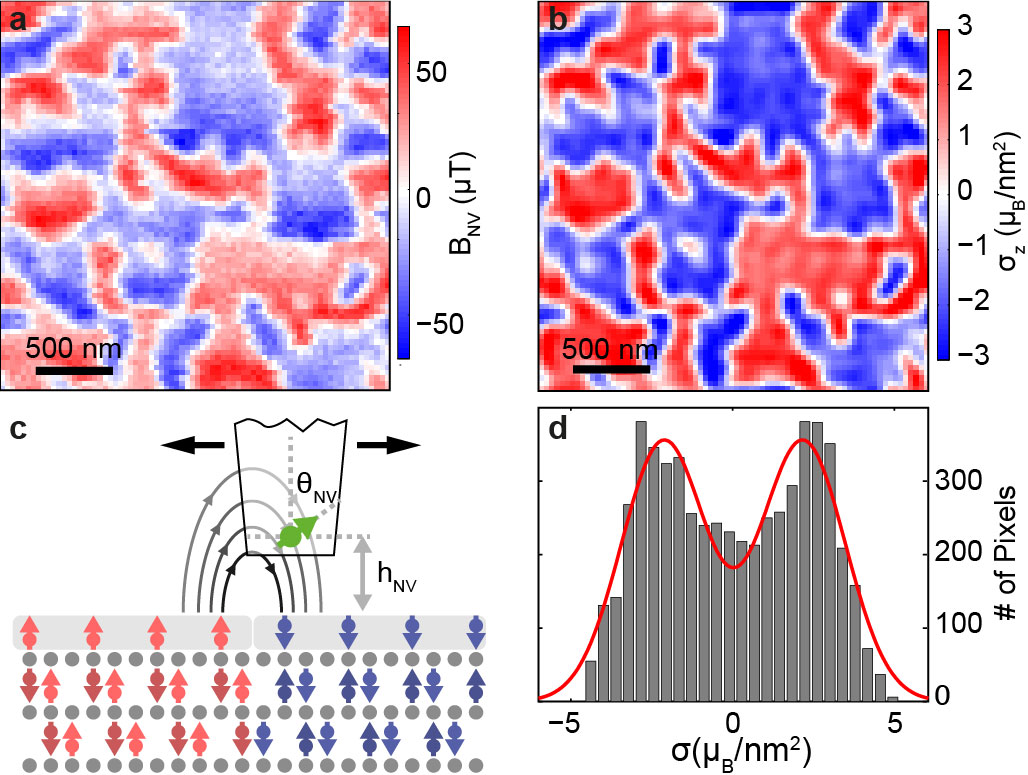}
\caption{{\bf Domain imaging in antiferromagnetic \CrO{}.} {\bf a} Map of the measured stray magnetic field $B_{\rm NV}$ above the \CrO{} film and {\bf b} the extracted moment density profile (see main text) of the film, which reveals a domain pattern of spin-up and spin-down domains. {\bf c} Measurement geometry and relevant experimental parameters for reverse propagation. {\bf d} Histogram of surface moment density values found in b with a fit to a bimodal, Gaussian distribution (red), yielding an average moment density $2.14\pm1.5~\mu_{\rm{B}}/$nm$^2$.
\label{Fig:Fig2}}
\end{figure}

To investigate AF domains in the \CrO{} film, we first use NV magnetometry to
acquire a map of $B_{\rm NV}$ at the \CrO{} surface, as shown in Fig.\,\ref{Fig:Fig2}a.  
The stray field image immediately confirms the presence of magnetic domains, as signalled by areas of positive or negative $B_{\rm NV}$. 
As expected from a simplified picture of homogeneously magnetised domains with infinitely sharp domain walls\,\cite{Tetienne2014}, zeros in $B_{\rm NV}$ indicate the locations of domain walls up to a small, constant shift, and broad maxima of $B_{\rm NV}$ occur towards the boundaries of the domains.
More quantitatively, we model this stray field by describing the magnetisation $\vec{m}$ of the \CrO{} film as two monolayers of out-of-plane polarised spins with moment density $\sigma_z(x,y)$ and opposite orientations, separated by $d=200~$nm (Fig.\,\ref{Fig:Fig1}b): %XXX
$\vec{m}(x,y,z) = \sigma_z(x,y)\left[\delta(z) - \delta(z+d)\right]\hat{z}$, where $\delta$ is the Dirac delta function and $\hat{z}$ the out-of-plane unit vector. The measured stray magnetic field $B_{\rm NV}$ can then be conveniently obtained by established methods of field propagation in Fourier space\,\cite{Meyer2004},
\begin{equation}\label{eq:BNV}
	B_{\rm NV}(\vec{k}) = \vec{m}(\vec{k})T_{\rm NV}(h_{\rm NV},\theta_{\rm NV},\phi_{\rm NV},\vec{k}),
\end{equation}
where $T_{\rm{\rm NV}}$ is a propagator that depends on the NV orientation $(\theta_{\rm NV},\phi_{\rm NV})$ and the NV-to-sample distance $h_{\rm NV}$ (see Fig.\,\ref{Fig:Fig2}c).
If $h_{\rm NV}$, $\theta_{\rm NV}$, and $\phi_{\rm NV}$ are known, Eq.\,(\ref{eq:BNV}) can be inverted and $\sigma_z(x,y)$ directly obtained from the experimental data together with appropriate filtering\,\cite{Meyer2004}. 
To determine these parameters, we developed an iterative, self-consistent method (see Appendix) based on the data and our minimal model for \CrO{}'s surface magnetisation described above. Using the resulting values $h_{\rm NV}=120~$nm, $\theta_{\rm NV}=54^\circ$, %XXX 
and $\phi_{\rm NV}=92^\circ$, we reverse propagate the measured $B_{\rm NV}(x,y)$ map to find $\sigma_z(x,y)$, as shown in Fig.\,\ref{Fig:Fig2}b.  This magnetisation profile shows well-defined magnetic domains, with typical domain sizes $\approx 230~$nm (determined by the peak width of the corresponding  autocorrelation map) and average surface moment densities 
$2.14\pm1.5~\mu_{\rm B}/$nm$^2$, as determined from a histogram of the inferred moment density map (Fig.\,\ref{Fig:Fig2}d). 
Independent confirmation of this value of $\sigma_z(x,y)$ was obtained through additional measurements on a patterned \CrO{} sample, which we prepared in a monodomain-state by field cooling (see Appendix).
The agreement between our experimental findings and the previous knowledge of surface magnetisation and moment densities in \CrO{}\,\cite{He2010,Dzyaloshinskii1992} 
supports the conclusion that the observed surface magnetisation indeed offers a faithful representation of the underlying AF domain pattern.

%TC:ignore 
%TC:endignore 

\begin{figure}
\includegraphics[width =8.6cm]{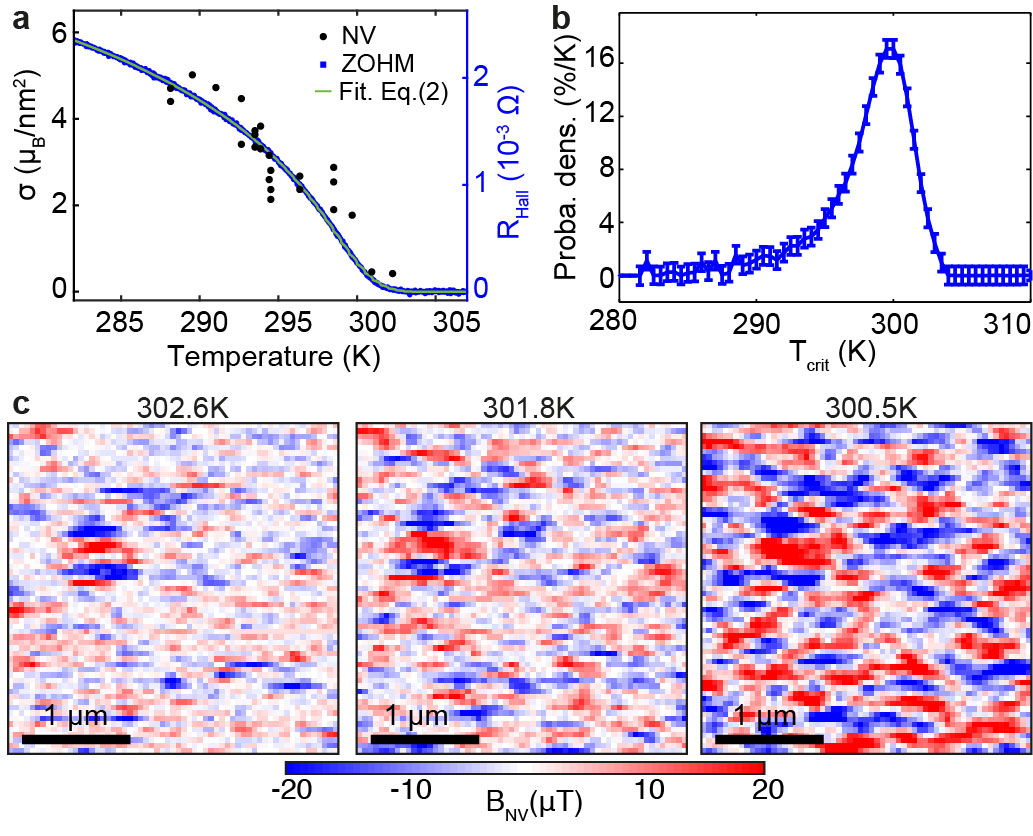}
\caption{{\bf Spatial variation of critical temperatures.} 
{\bf a} Temperature dependence of moment density together with 
zero-offset Hall magnetometry data\,\cite{Kosub2015}.
The data are fitted (green) to a critical behaviour with spatially varying critical temperatures (see eq.\,(\ref{EqSigAvg}) and text for details). 
{\bf b} Probability distribution of critical temperatures as determined from the fit in a. 
{\bf c} Consecutive magnetic field maps obtained during cooling of the sample through the phase transition. Domains are nucleating at isolated spots and maintain their orientation during the subsequent formation of the domain pattern. \label{Fig:Fig3}}
\end{figure}

Further details on the nanoscale magnetic properties of our thin film AF can be obtained by observing the temperature dependence of $\sigma_z$ near the AF-paramagnet transition at the critical temperature of the thin film.   
We thus repeated moment density measurements for temperatures around $T_{\rm Neel}$  
and compared our findings to zero-offset Hall magnetometry (ZOHM) measurements performed on a similarly prepared sample\,\cite{Kosub2015} (Fig.\,\ref{Fig:Fig3}a).  Briefly, ZOHM measures the anomalous Hall resistance in a thin layer of Pt evaporated onto the \CrO{} surface, and is sensitive to the average \CrO{} surface magnetic moment over the electrode area of $\sim900\,\mu$m$^2$.  This method gives a precise readout of the averaged, relative magnetisation, but does not provide spatial resolution or the magnitude of the moment density.  Clearly, our NV magnetometry data (Fig.\,\ref{Fig:Fig3}a) follow the temperature dependence of ZOHM and thereby allow us to calibrate the resulting Hall resistance to a quantitative magnetic moment density.

The temperature dependance of $\sigma_z$ determined by combining ZOHM and NV magnetometry (Fig.\,\ref{Fig:Fig3}a) shows a smooth tapering of $\sigma_z$ through the phase transition, in contrast to the sharp drop to zero expected from the usually observed power-law dependance of $\sigma(T)$  for magnetic phase transitions.
Such behaviour was previously attributed to spatial variations of $\Tsub{crit}$ in thin films\,\cite{Fallarino2015,Muftah2016}  
and can be readily accounted for by the convolution
\begin{equation}  \label{EqSigAvg}
\sigma_{\rm{avg}}(T) = \int P(T_{\rm{crit}})\sigma\left(\frac{T}{T_{\rm{crit}}}\right) dT_{\rm{crit}},
\end{equation}
where $P(\Tsub{crit})$ is the probability density for $\Tsub{crit}$ and 
$\sigma(\tau) = \sigma_{\rm sat}\left(1-\tau\right)^\beta$, 
with  critical exponent $\beta$ and saturation magnetisation $\sigma_{\rm sat}$ ($\sigma(\tau>1)=0$).
Fitting Eq.\,(\ref{EqSigAvg}) to our data (green curve in Fig.\,\ref{Fig:Fig3}a) for fixed $\beta=0.35$ 
allows us to extract $P(\Tsub{crit})$ as depicted in Fig.\,\ref{Fig:Fig3}b. The significant broadening of $P(\Tsub{crit})$ is evidence of local inhomogeneity of $\Tsub{crit}$ in the sample, which we assign to material defects such as twinning boundaries or lattice dislocations\,\cite{Kosub2017}.

%TC:ignore 
%TC:endignore 

Our ability to measure fine spatial features of the magnetisation pattern enables a detailed study of the paramagnetic-to-AF phase transition and the local variations in $\Tsub{crit}$ at the level of individual domains.  To directly observe these variations, we record snapshots of the stray field  
around $\Tsub{crit}$
(Fig.\,\ref{Fig:Fig3}c).  For the sample well above $\Tsub{crit}$, we detect no magnetic stray field exceeding our measurement noise, which indicates no surface magnetisation and a paramagnetic phase.  As the sample is cooled, spatially separated regions of non-zero magnetisation spontaneously nucleate, but significant areas of the sample remain paramagnetic. 
Upon further cooling, i.e. below $300.5~$K (Fig.\,\ref{Fig:Fig3}c, right), this nucleation propagates until all areas of the sample show non-zero magnetisation. 
This lateral spreading of AF domains with decreasing temperature is indicative of significant, inter-granular exchange coupling. Without such coupling, each grain would nucleate independently and in a random fashion and no domains larger than the grain size would occur. 

\begin{figure}
\includegraphics[width=8.6cm]{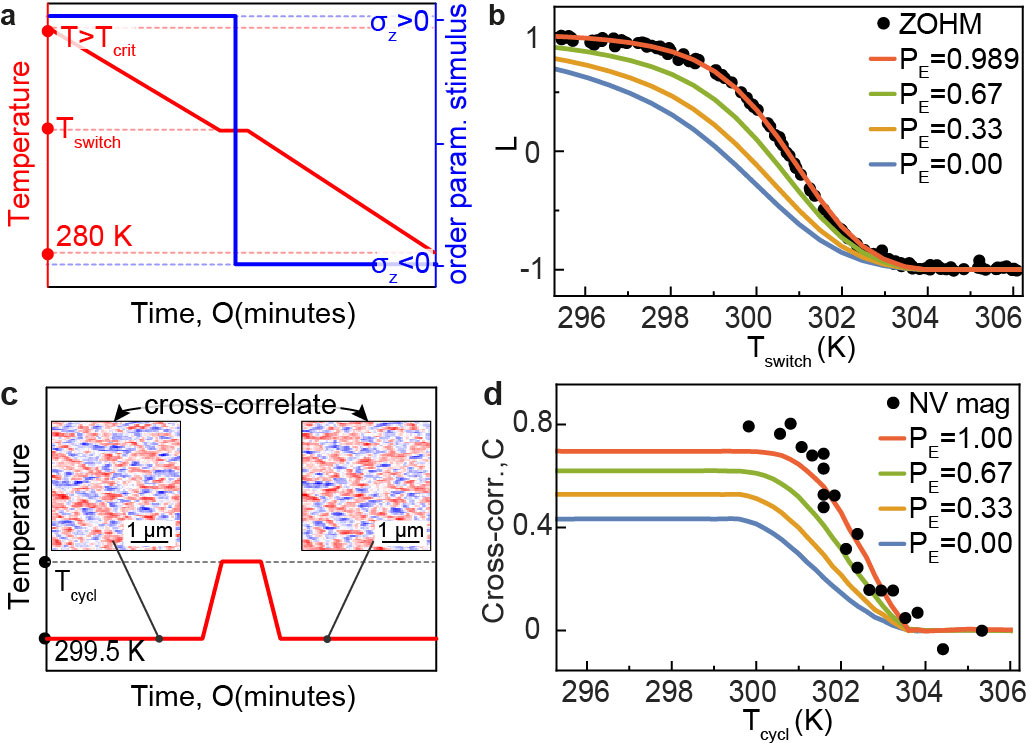}
\caption{{\bf Determining inter-granular exchange coupling}. {\bf a} Procedure for differential field cooling. The sample was cooled from a temperature $T>T_{\rm{crit}}$ to $T=T_{\rm{switch}}$ while applying a strong order parameter selection stimulus (magnetic field) towards $\sigma_z>0$. The stimulus was then inverted towards $\sigma_z<0$ and the sample further cooled to $280~$K. 
{\bf b} The average order parameter $L=\left<\sigma_z(\Tsub{switch})\right>/\left<\sigma_z(\Tsub{RT})\right>$, measured by ZOHM, as a function of $T_{\rm{switch}}$. 
Data are fit to theory (see text) using the probability distribution from Fig.\,\ref{Fig:Fig3}b and demonstrate significant inter-granular exchange coupling $P_{\rm{E}}\sim1$. 
{\bf c} Procedure for temperature cycling domain cross-correlation.  The sample is imaged at $T_{\rm start}=299.5~$K, subsequently heated to $T=T_{\mathrm{cycl}}$ and cooled back to $T_{\rm start}$, where another image is taken. 
{\bf d} Cross-correlation between reference and sample image as a function of $T_{\mathrm{cycl}}$, evidencing strong inter-granular exchange coupling on the nanoscale. \label{Fig:Fig4}} 
\end{figure}

In order to test this interpretation, 
we developed a differential field cooling (DFC) method (Fig.\,\ref{Fig:Fig4}a), which allows us to assess the efficiency of inter-granular exchange coupling through ZOHM\,\cite{Kosub2017b}. 
We initialised the sample well above $\Tsub{crit}$, and then applied a strong magnetic field in the $+\hat{z}$ direction to provide an order parameter selection stimulus\,\cite{Kosub2017} towards an AF state with $\sigma_z>0$, while the sample is cooled. 
At a temperature $\Tsub{switch}$, the magnetic field is reversed to yield a stimulus towards $\sigma_z<0$, with which the sample is further cooled to $280~$K. Finally we measure the average order parameter $\left<\sigma_z(\Tsub{switch})\right>$ using ZOHM and determine the normalised magnetisation $L=\left<\sigma_z(\Tsub{switch})\right>/\left<\sigma_z(\Tsub{RT})\right>$ as a function of $\Tsub{switch}$ (Fig.\,\ref{Fig:Fig4}b).
To understand the evolution of $L$, we developed a discrete model of the AF film (see Appendix), taking into account inhomogeneities in $\Tsub{crit}$ and $\sigma_z$. 
The discrete grains comprising the film are mutually exchange coupled with a probability $P_E$ for the order-parameter of neighbouring grains to be locked by exchange (i.e. for exchange coupling exceeding thermal fluctuations).   
For $P_E=0$, grains are independent and their value of $\sigma_z$ determined by the applied selection stimulus at the time of ordering.
Therefore, $L(\Tsub{switch}) = -1 + 2 A(\Tsub{switch})$, with $A(\Tsub{switch})=\int_{\Tsub{switch}}^{\infty} P(T_{\rm crit})\,dT_{\rm crit}$ the unit-less fractional sample area having $\Tsub{crit}>\Tsub{switch}$.  
For $P_E>0$, however, each grain will be influenced by its neighbours and we find $L(\Tsub{switch},P_E)= -1 + 2 A(\Tsub{switch})e^{P_E\cdot(1-A(\Tsub{switch}))}$ (see Appendix).  
A fit of $L(\Tsub{switch},P_E)$ to the ZOHM data (Fig.\,\ref{Fig:Fig4}b) thus allows us to determine $P_E=\,0.989$, indicating that for our thin film sample, inter-granular exchange coupling largely dominates thermal fluctuations.

%TC:ignore 
%TC:endignore 

Scanning NV magnetometry confirms such strong exchange coupling of neighbouring grains on the level of individual AF domains.  
For this, we perform domain cross-correlation measurements under temperature cycling (Fig.\,\ref{Fig:Fig4}c). 
We record an AF domain image at an initial temperature $T_{\rm start}=299.5~$K, subsequently set the sample temperature to $\Tsub{cycl}$ for $\sim10\,$min, and finally cool the sample back to $T_{\rm start}$, where we record a second domain image. This procedure yields information closely related to DFC, all while avoiding experimental difficulties of applying and switching strong order parameter stimuli (i.e. magnetic fields) in our imaging experiment. Specifically, the pixel-by-pixel correlation $C(T_{\rm cycl})$ between reference and sample images 
(Fig.\,\ref{Fig:Fig4}d) measures the extent to which the domain pattern is preserved during heating. 
In analogy to DFC, we expect the correlation between grains to follow $A(T_{cycl})$ for $P_E=0$, while for $P_E>0$ we expect the domain pattern to be largely determined by high-$T_{crit}$ grains, and the transition from correlated to uncorrelated to therefore be pushed to higher temperatures, as confirmed by granular thin-film simulations (see Appendix).
Comparison of our correlation data (Fig.\,\ref{Fig:Fig4}d, black points) to the simulation (solid lines) provides further proof of the strong inter-granular exchange coupling, which we obtain on the level of individual domains. 

In this work, we have demonstrated a novel, versatile technique for quantitative, nanoscale imaging of AF order in thin film materials. Our approach is largely  complementary to the existing toolset for studying AF order such as neutron\,\cite{Shull1949,Sosnowska1982,Schreyer2000}, or X-ray \,\cite{Nolting2000,Alders1998,Kuiper1993,Mertins2001,Valencia2010} scattering as well as optical\,\cite{Saidl2017}, or scanning probe imaging\,\cite{Wu2011}, where combining high spatial resolution with high signal-to-noise ratio and quantitative imaging has remained elusive thus far.
Importantly, our approach is not restricted to \CrO{}. Indeed, the occurrence of boundary magnetisations in AFs is related to inversion symmetry breaking on the surface\,\cite{Andreev1996, Dzyaloshinskii1992}. NV magnetometry is therefore generally applicable to AFs, as long as the resulting surface spin densities are within our sensitivity limit. 
Our results thus constitute an important step towards the general understanding of surface magnetisation and domain formation in AF's 
and firmly establish scanning NV magnetometry as a versatile tool for further developments of nanoscale AF spintronic technologies\,\cite{Jungwirth2016}.

\begin{acknowledgments}
We thank Prof.\,O.\,G.\,Schmidt (IFW Dresden) for his insightful input at the initial stage of the project and S.\,Hoffman (Basel) and V.\,Jacques (CNRS Montpellier) for helpful discussions.  We gratefully acknowledge financial support through the NCCR QSIT, a competence center funded by the Swiss NSF, the SNI and through SNF Grant No.~143697 and 155845. This research has been partially funded by the European Commission's~7.\ Framework Program (FP7/2007-2013) under grant agreement number 611143 (DIADEMS), ERC within the EU 7th Framework Programme (ERC Grant No.~306277) and the EU FET Programme (FET Open Grant No.~618083).
\end{acknowledgments}

\appendix

\section{NV magnetometry}

Scanning NV magnetometry\,\cite{Rondin2014} was performed under ambient conditions, using single-crystal diamond scanning probes as described in\,\cite{Appel2016}. For the measurements, a single NV centre contained within the tip was scanned within $\sim100~$nm of the \CrO{} surface. For quantitative magnetometry, the electron spin resonance (ESR) frequency was recorded by locking a microwave driving field to the ESR transition \cite{Schoenfeld2011}.
For this, a green laser (power level $\sim 100~\mu$W) was used to excite NV fluorescence, resulting in typical fluorescence count rates $\sim700~$kHz and ESR contrasts $\sim15~\%$. In order to obtain a sign-sensitive measurement of $B_{\rm NV}$, a bias magnetic field of $24~$G was applied along the NV axis during all measurements.  The single-pixel integration time for the measurements presented here ranges from $0.6~$s (temperature cycling correlation images used in Fig.\,4c,d) to $7~$s (Fig.\,2a). 

\section{Sample fabrication}

The \CrO{} films were grown by reactive evaporation of chromium from a Knudsen cell in high vacuum onto $c$-cut sapphire substrates (Crystec GmbH) heated to $700~^\circ$C initially and to $500~^\circ$C after the first few monolayers. The background gas used was molecular oxygen at a partial pressure of $10^{-5}~$mbar. The deposition was carried out using rates of about $0.04~$nm\,s$^{-1}$ and was monitored in situ by reflection high-energy electron diffraction. \CrO{} layers were subjected to a vacuum annealing process at $750~^\circ$C and residual pressure of $10^{-7}~$mbar directly after growth. The thin Pt top layers were magnetron-sputter-deposited at lower temperatures of $\approx100~^\circ$C using a higher rate of $0.1~$nm\,s$^{-1}$ to maintain layer continuity. Hall crosses were patterned from the top Pt layers, by SF$_6$ reactive ion etching around a photoresist mask.

\section{Determination of $\sigma(x,y)$ from 2-D magnetic stray field profiles}

We determine the surface moment profile $\sigma(x,y)$ (Fig.\,2b,d) from our measurement of the magnetic field $B_{\rm{NV}}(x,y,h_{\rm{NV}})$ in the plane at a height $h_{\rm{NV}}$ above the surface of the film (Fig.\,2a) via a Fourier propagation method.\cite{Meyer2004}  We assume a perpendicularly magnetised layer of spins of moment density $\sigma(x,y)$ at the top surface of the film ($z=0$), and an oppositely magnetised layer at the bottom surface of the film ($z=-d$, with film thickness $d$).  It is convenient to work in Fourier space, where we consider $\tilde{\sigma}(k_x,k_y)$, the two-dimensional Fourier transform of $\sigma(x,y)$ (we will continue to work in real space along the $z$-axis).  Because there are no external time-varying electric fields or currents, we can define a magnetic scalar potential $\tilde{\phi}(k_x,k_y,z)$ that is the solution to the Laplace equation with boundary conditions set by $\tilde{\sigma}(k_x,k_y)$.  Consider first the top layer of spins at $z=0$. The potential at a height $z$ is found from the moment density by $\tilde{\phi}_{top}(k_x,k_y,z) = \tilde{\phi}_{top}(k_x,k_y,0)e^{-kz} = \tilde{\sigma}(k_x,k_y)e^{-kz}/2$, where $k=(k_x^2+k_y^2)^{1/2}$.  The magnetic field is then computed from the gradient of $\tilde{\phi}_{top}(k_x,k_y,z)$.  Adding the contribution from the bottom layer of spins we thus have the magnetic field in the plane of the NV, $\tilde{\vec{B}}(k_x,k_y,h_{\rm{NV}}) = -\mu_0\vec{\nabla}_{k}(\tilde{\phi}_{top}+\tilde{\phi}_{bottom})$, with in-plane and perpendicular components given by
\begin{eqnarray}
&\tilde{B}_{x,y}(k_x,k_y,h_{\rm{NV}}) =  -i\mu_0 k_{x,y} \frac{e^{-kh_{\rm{NV}}}-e^{-k(h_{\rm{NV}}+d)}}{2}\tilde{\sigma}(k_x,k_y) \equiv T_{x,y}\tilde{\sigma}(k_x,k_y)\\
&\tilde{B}_{z}(k_x,k_y,h_{\rm{NV}}) =  \mu_0 k \frac{e^{-kh_{\rm{NV}}}-e^{-k(h_{\rm{NV}}+d)}}{2}\tilde{\sigma}(k_x,k_y) \equiv T_z\tilde{\sigma}(k_x,k_y).
\end{eqnarray}
We now relate the field projection along the NV axis, $\tilde{B}_{\rm{NV}}(k_x,k_y,h_{\rm{NV}})$, to the surface moment density through a single propagator $T_{\rm{NV}}$:
\begin{eqnarray}
\tilde{B}_{\rm{NV}}(k_x,k_y,h_{\rm{NV}}) &=& \sin(\theta_{\rm{NV}})\cos(\phi_{\rm{NV}})\tilde{B}_x + \sin(\theta_{\rm{NV}})\sin(\phi_{\rm{NV}})\tilde{B}_y + \cos(\theta_{\rm{NV}})\tilde{B}_z \\
&\equiv& T_{\rm{NV}}\tilde{\sigma}(k_x,k_y)
\end{eqnarray}

Using $T_{\rm{NV}}$, the moment density profile can be transformed into a magnetic field at a height $z$ from the sample. Moreover, using the inverse propagator a reverse propagation can be performed and the moment density profile can be directly determined from the 2 dimensional field map measured at a distance $h_{\rm{NV}}$ above the sample. 

High frequency components (large $k$) of the moment density get damped by the exponential factor in the propagator. Conversely, by performing the reverse propagation, high frequency oscillations (including measurement noise) get enhanced. 
To filter out such higher frequency noise, we introduce a filter function given by a Hanning window \cite{Roth1989}
\begin{equation}  \label{eq:5}
W(k)=
\begin{cases}
0.5[1+\cos(\pi \left[  kh_{\rm{NV}}/2\pi\right] )],& \text{for }  kh_{\rm{NV}}/2\pi<1\\
0,              & \text{for }  kh_{\rm{NV}}/2\pi>1
\end{cases}
\end{equation}
The cutoff frequency of $k_{\rm{cutoff}}=2\pi/h_{\rm{NV}}$ is motivated by the fact that the NV centre can only resolve oscillations with a frequency given by the NV to sample distance $h_{\rm{NV}}$.

The magnetic moment density profile can finally be determined from the measured magnetic field map using
\begin{equation} \label{Eq:backprop} 
\sigma(k) = T_{\rm{NV}}^{-1}(h_{\rm{NV}},\theta_{\rm{NV}},\phi_{\rm{NV}}) W(k) B_{\rm{NV}}(k). 
\end{equation}

In order to apply this equation to find the underlying moment density profile, we apply the following procedure.  First, we reverse propagate the field using an approximate starting value for $h_{NV}$, $\theta_{\rm{NV}}$ and $\phi_{\rm{NV}}$ to find a moment density $\sigma(x,y)$.  From $\sigma(x,y)$ we find the domain boundaries, and assume a uniform moment density within a given domain, $\sigma_{\pm}=\sigma_0\, \mathrm{sign}[\sigma(x,y)]$.  We then forward propagate $\sigma_{\pm}(x,y)$ and compare with the original stray field data.  Using a least square fitting routine, we then find the values of the NV-to-sample distance\,($h_{\rm{NV}}=120\,$nm), the NV orientation ($\theta_{\rm{NV}}=54\,^\circ$, $\varphi_{\rm{NV}}=92\,^\circ$) and the moment density ($\sigma_0=3\,\mu_{\rm{Bohr}}/$nm$^2$) that best reproduce the measured magnetic field. The fitted values of $h_{NV}$, $\theta_{\rm{NV}}$ and $\phi_{\rm{NV}}$ are then used for a final back-propagation to generate the moment density profile depicted in Fig.\,2b.

\section{Surface moment density determined via  uniformly magnetised stripes of \CrO{}}

To confirm the measurement of surface moment density obtained by Fourier propagating the magnetic field of the domain pattern, we additionally patterned a stripe into the \CrO{} film and imaged the stray field of a uniform magnetized film, which we generated by magnetic field cooling. 

We structured 1-$\mu$m-wide Cr$_2$O$_3$ stripes from a 200-nm-thick \CrO{} film coated with $2\,$nm of Pt. We first fabricated stripe masks using ebeam lithography ($30\,$keV) with a hydrogen silsesquioxane ebeam resist (HSQ, FOX-16 Dow Corning). The mask was afterwards transferred into the \CrO{} using inductively coupled plasma reactive ion etching (ICP-RIE, Sentech SI 500) for $120\,$s in an ArCl$_2$ plasma ($40\,$sccm Cl$_2$, $25\,$sccm Ar, $1.0\,$Pa pressure, $400\,$W ICP power, $100\,$W RF power with $-232\,$V bias). We removed the etch mask using buffered oxide etch (10:10:1 deionized water, ammonium fluoride,  $40\,\%$ HF) for $60\,$s. The etch depth was $250\,$nm, ensuring that the unmasked \CrO{} film is fully etched through. We also observed a tapering of the sidewalls, which we later included in the stray field calculation.

We model the stray field of a uniformly magnetized \CrO{} film as a stack of two oppositely magnetized ferromagnetic layers, separated by \SI{200}{\nm}, and determine the moment density by a fit of our data to this model.
A ferromagnetic film with out of plane anisotropy can be seen as the magnetic counterpart of planar capacitor.\cite{Hingant2015} For a thin film, this results in a stray field that can be described by a current at the edge of the film
\begin{equation}  \label{eq:B_Stripex1}
B_x(x,h_{\rm{NV}})= \frac{\mu_0 I}{2\pi}\frac{h_{\rm{NV}}}{h_{\rm{NV}}^2+x^2}
\end{equation}

\begin{equation}  \label{eq:B_Stripez1}
B_z(x,h_{\rm{NV}})= \frac{\mu_0 I}{2\pi}\frac{x}{h_{\rm{NV}}^2+x^2},
\end{equation}
where the current $I$ is set by the moment density.

\begin{figure}[!htb]
	\includegraphics[width=1\columnwidth]{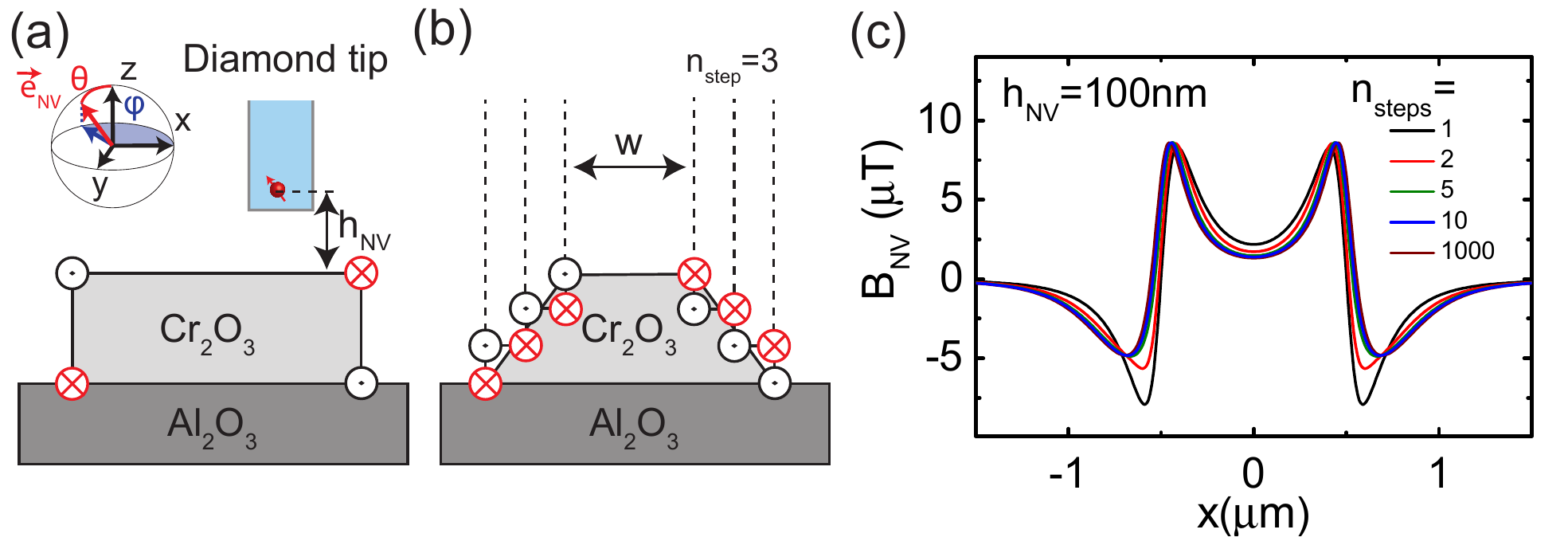}
	\caption[Schematic of a \CrO{} stripe]{(a) Schematic of a \CrO{} stripe. 
			The magnetic field of the stripe can be calculated as the magnetic field of four currents at the edges. (b) The sidewalls of a tapered stripe are approximated as stairs carrying currents at the edges. (c) Calculated magnetic field of a stripe (width $w=1\,\mu$m) with a tapered sidewalls ($45^\circ$) in a height $h_{\rm{NV}}=100\,$nm ($\theta_{\rm{NV}}=54^\circ$, $\varphi_{\rm{NV}}=90$). The stray field for different numbers of steps indicates the convergence of this model.   
		\label{Fig:stripemodel}}
\end{figure}

For the AF film, we assume two such thin, oppositely magnetised films, one at the top surface and the other at the bottom surface. The edge of an AF film is therefore described by an edge-current at the top and a second such current at the bottom edge, running in opposite directions (see Fig.\,\ref{Fig:stripemodel}a). Thus, the stray field is given by $B_{x,z}^{\rm{edge}}=B_{x,z}(x,h_{\rm{NV}})-B_{x,z}(x,h_{\rm{NV}}+d)$. The measured AF stripes consist of two edges and the magnetic stray field can be interpreted as the magnetic field of four currents at the edges, as illustrated in Fig.\,\ref{Fig:stripemodel}a. The stray field is then given by $B_{x,z}^{\rm{stripe}}=B^{\rm{edge}}_{x,z}(x-w/2,h_{\rm{NV}})-B^{\rm{edge}}_{x,z}(x+w/2,h_{\rm{NV}})$, as depicted in Fig.\,\ref{Fig:stripemodel}c (black curve).

The etch did not produce a vertical etch but left tapered sidewalls to the stripe. We take this tapering of the sidewall into account by decomposing it into $n_{\rm{steps}}$ steps (see Fig.\,\ref{Fig:stripemodel}b) and leaving the angle of the sidewall as a fit parameter. The final magnetic field is calculated as a field produced by $4n_{\rm{steps}}$ currents. The stray field for an angle of $45^\circ$ is plotted in Fig.\,\ref{Fig:stripemodel}c for different $n_{\rm{steps}}$, illustrating the convergence of the model. Finally, we fitted this model (where we chose $n_{\rm{steps}}=10$ as a trade-off between calculation time and accuracy of the model), to the measured magnetic field map to precisely determine the NV-to-sample distance and the moment density.  

\begin{figure}[!htb]
	\includegraphics[width=1.00\columnwidth]{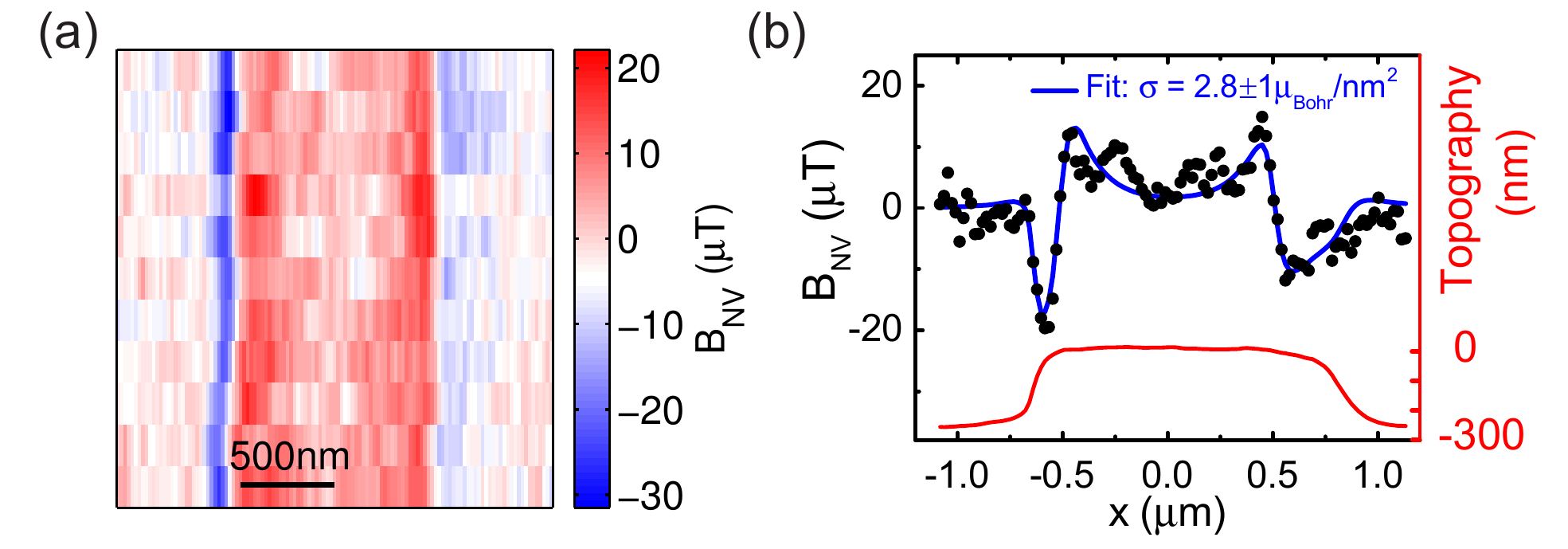}
	\caption[Determination of the surface moment density]{(a) Measured stray field map of the uniformly magnetized stripe. (b) Linecut of the magnetic field across the stripe. Black dots corresponds to the measurement and the blue line is the fitted field profile. The fitting of 16 linecuts yields an NV-to-sample distance $h_{\rm{NV}}=89\pm23\,$nm and a moment density of $\sigma = 2.8\pm1\, \mu_{\rm{Bohr}}/$nm$^2$.
		\label{Fig:momdens}}
\end{figure}

Fitting this model to the stray field of $16$ lines perpendicular to the stripes (Fig.\,\ref{Fig:momdens}) yields a distance of $h_{\rm{NV}}=89\pm23\,$nm and a moment density of $\sigma = 2.8\pm0.7\, \mu_{\rm{Bohr}}/$nm$^2$. This value confirms the moment density determined via the back-propagation method shown in Fig.\,2 in the main text and outlined above. 

\section{Determination of $P(T_{\rm{crit}})$}

The ZOHM measurement of $R_{\rm{Hall}}(T)$ (Fig.\,3a) indicates the presence of a wide range of critical temperatures in our film.  As described in the main text, we determine the distribution of critical temperatures, $P(T_{\rm{crit}})$, from a convolution of the single-critical-temperature function $\sigma(\tau)=(1-\tau)^{0.35}\Theta(1-\tau)$, where $\Theta$ is the Heaviside step function.  In practice, we divide our temperature range into a set of discrete, \SI{0.5}{\kelvin}-wide bins to construct the fit function $R_{\rm{Hall}}(T)= \sum_i p_i\sigma(T/T_i)$, and find the fit parameters $\{p_i\}$ that best describe our data.  The resulting histogram is shown in Fig.\,3b.

\section{Temperature calibration}

The measurements reported here were taken with 3 different temperature controlled apparatuses, which must be calibrated to each other.  The ZOHM measurement of $R_{\rm{Hall}}(T)$ (Fig.\,3a) is taken as the reference measurement, from which the critical temperature distribution $P(T_{\rm{crit}})$ is determined, as above.  The DFC measurement (Fig.\,4a-b) was taken with a different peltier heater and temperature sensor, so the temperature is shifted in order to fit the data to the theoretical curve (which is derived from $P(T_{\rm{crit}})$).  Finally, all of the NV magnetometry data were taken with a third peltier stage and thermistor, which we also calibrate to the ZOHM measurement of $R_{\rm{Hall}}(T)$.  To do so, we use the NV magnetometry measurement of domain cross-correlation under temperature cycling (Fig.\,4d), which has a sharper temperature dependence than the measurement of magnetization vs.~temperature (Fig.\,3a).  We thus shifted the temperature of our NV magnetometry measurements by \SI{+0.5}{\kelvin}, so that the turn-on temperature for the cross correlation signal aligns with the simulation based on $P(T_{\rm{crit}})$ (Fig.\,4d).

\section{Granular model of ordering dynamics}

To model the ordering dynamics of our thin film, we introduce a framework based on an ensemble description of fundamental indivisible AF entities, which we attribute to crystallographic grains.\cite{Kosub2017b}  The order parameter within a given grain is uniform and is determined by a combination of external fields and exchange coupling to neighbouring grains.  Because the size of a grain defines its magnetic anisotropy energy, and therefore its thermal stability, the critical temperature at which a grain undergoes the PM-AF phase transition is size dependent and falls within the distribution shown in Fig.\,3b, with the largest grains having the highest $T_{\rm{crit}}$.  As the film is cooled through this temperature distribution, the grains therefore order sequentially by size, with the order parameter $L$ of each grain taking a value of $\pm 1$.  We describe the ordering outcome for each grain probabilistically as detailed below, with a single phenomenological quantity $P_E$ parameterising the strength of the exchange coupling between grains.  By simulating $\langle L \rangle$ for a sufficiently large sample size, we can model the expected behaviour of our film in both the DFC and domain cross-correlation measurements.

\subsection{Methodology}

\begin{figure}[!htb]
	\includegraphics[width=1.00\columnwidth]{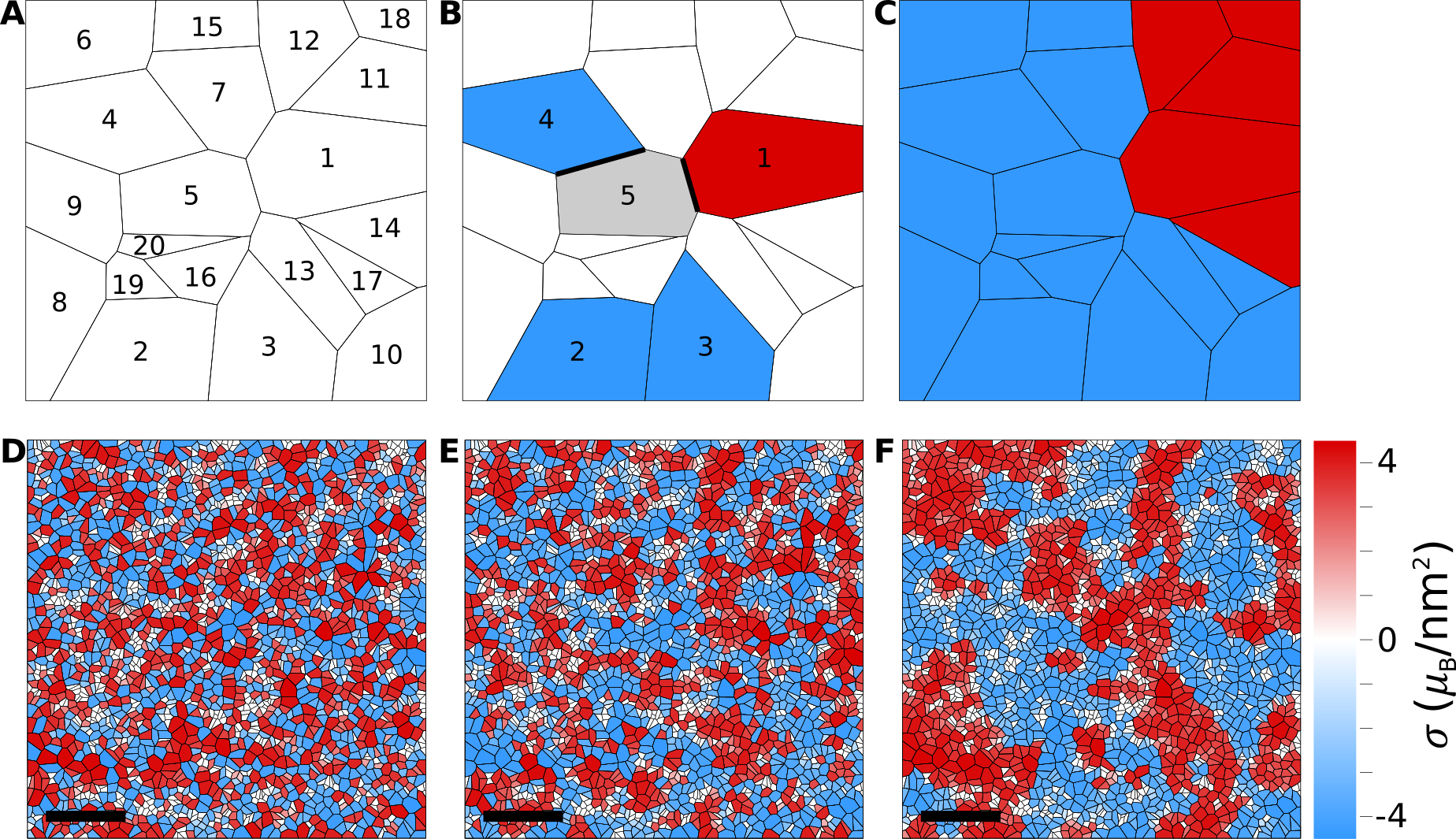}
	\caption {\textbf{Magnetic ordering of grains.} (a-c) Ordering of 20-grain mesh for $P_E=1$, $\eta=0$.
	(a) Voronoi mesh based on 20 randomly selected points, with grains are ordered by area from largest to smallest.  Above $T_{\rm{crit}}$ all grains are in a paramagnetic state.
	(b) After some cooling, grains 1-4 have formed 3 distinct nucleation spots, with $L^{1,2,4}$ selected randomly, while $L^3$ is determined from exchange coupling to grain 2.  As the temperature is lowered through $T_{\rm{crit}}^5$, $L^5$ is determined from the average of $L^1$ and $L^4$, weighted by border length (thick black lines).
	(c) After further cooling, the sample has completely transitioned to the AFM state, and the 3 initial nucleation spots have spread to form 2 domains.
	(d-f) Simulation of zero-field cooling on a random mesh with (d) $P_E=0.0$, (e) $P_E=0.5$, and (f) $P_E=1.0$, after cooling to a temperature of \SI{294.15}{\kelvin}.  In the case of strong exchange coupling between grains, the zero-field cooled magnetisation pattern closely resembles the domain images observed in the experiment, whereas for weaker exchange coupling the simulated domains are more fragmented.  (Scale bars: \SI{0.5}{\micro\meter}).
\label{Fig:mesh}}
\end{figure}

Given a set of grains $\{i\}$ ordered by inverse critical temperature $\{T_{\rm{crit}}^i\}$, and having areas $\{A^i\}$, we determine the order parameter $L^{i}$ of grain $i$ based on two possible influences, as follows.  First, if grain $i$ has antiferromagnetically ordered neighbours, then with probability $P_E$, $L^{i}$ is determined by exchange coupling to neighbouring grains, in which case $L^{i}$ is selected randomly, such that the expectation value of $L^{i}$ is determined by the average order parameter of the neighbouring grains.  Alternatively, with probability $1-P_E$, or if all neighbouring grains are still paramagnetic, $L^i$ is not determined by exchange coupling.  In this case, $L^i$ is assigned to $+1$ with probability $(\eta^i+1)/2$, and to $-1$ otherwise.  The introduction of $\eta^i$ allows for the possibility of an externally applied field that biases the ordering in one direction or the other.  $\eta^i$ can therefore take on any value in the range $[-1,+1]$, with $\eta^i=0$ corresponding to no applied field.  In the DFC experiment, $\eta$ is set to $+1$ initially and is flipped to $-1$ at temperature $T_{\rm{switch}}$.  In the domain cross-correlation experiment, $\eta=0$ always.  An example of cooling 20 grains with $P_E=1$ and $\eta=0$ is shown in Fig. \ref{Fig:mesh}(a-c).

With the resulting $\{L^i\}$ and $\{T_{\rm{crit}}^i\}$ from such a simulation, we calculate the observable quantities $\langle L \rangle$ and $B_{\rm{NV}}(x,y,h_{\rm{NV}})$.

\subsection{Differential Field Cooling}

In the differential field cooling experiment, the external field is initially set to a high value ($\eta_{\rm{high}} = 1$) and then is reversed at $T_{\rm{switch}}$ ($\eta(T>T_{\rm{switch}}) = \eta_{\rm{low}} = -1$).  Measuring the anomalous Hall resistance over the area of the Hall cross yields the average order parameter $\langle L \rangle$.  We model this process on set of grains $\{i\}$, as outlined above, but do not make any assumption about the spatial layout of the grains.  Instead, we use the macroscopic properties of the film, namely the fractional ordered area and the average order parameter at the time of ordering of grain $i$, as a statistical average description of the neighbours of grain $i$.  Thus, the order parameter of grain $i$ becomes:
\begin{equation}
	\langle L^i \rangle = P_E\frac{\sum_{j<i}L^jA^j}{A_{tot}} + \left[1-P_E\frac{\sum_{j<i}A^j}{A_{tot}}\right]\eta^i,
\end{equation}
where $A_{tot}$ is the total sampled area.  In the limit of large number of grains, the average order parameter is given by 
\begin{equation}\label{eq:DFCanalytic}
	\langle L\rangle(T_{\rm{switch}},P_E) = \eta_{\rm{low}} + (\eta_{\rm{high}}-\eta_{\rm{low}})\frac{A_{\rm{AF}}(T_{\rm{switch}})}{A_{\rm{tot}}}e^{P_E\left(1-\frac{A_{\rm{AF}}(T_{\rm{switch}})}{A_{\rm{tot}}}\right)},
\end{equation}
where $A_{\rm{AF}}(T_{\rm{switch}})$ is the ordered area at the time of switching:
\begin{equation}
	A_{\rm{AF}}(T_{\rm{switch}})=\int_{T_{\rm{switch}}}^{\infty} \tau_A(T_{\rm{crit}})\,dT_{\rm{crit}}.
\end{equation}
This result is shown by the solid curves in Fig.\,4b for several values of $P_E$.  In the limit of no exchange coupling, i.e. $P_E=0$, the order parameter of grain $i$ is determined by $\eta^i$, so that the average order parameter is found by simply by integrating over the ordered area at the time of switching:
\begin{equation}
	\langle L\rangle(T_{\rm{switch}},P_E=0) = \eta_{\rm{low}} + (\eta_{\rm{high}}-\eta_{\rm{low}})\frac{A_{\rm{AF}}(T_{\rm{switch}})}{A_{\rm{tot}}}.
\end{equation}
On the other hand, with the addition of exchange coupling between grains, the area that orders before the switch ($\eta_{\rm{high}}$) influences the ordering of subsequent grains, resulting in an effectively larger area experiencing the $\eta_{\rm{high}}$ stimulus.  The comparison in Fig.\,4b between the experimental results and the analytic expression in Eq.\,\ref{eq:DFCanalytic} indicates a high degree of exchange coupling between grains, with $P_E=0.989$.

\subsection{Domain cross-correlation under temperature cycling}

We further investigated the effects of exchange coupling microscopically via domain cross correlation under temperature cycling.  In this case, to fully simulate the experiment we require a real-space grain configuration, so we begin by defining a random Voronoi mesh that approximates the grain size distribution in our film.  Furthermore, we assign a critical temperature to each grain such that the area-weighted distribution of critical temperatures for the mesh follows the distribution in Fig.\,3b, and such that $T_{\rm{crit}}^i > T_{\rm{crit}}^{i+1}$.  Since we know exactly the state of each grain's neighbours, we can assign the order parameter of grain $i$ directly instead of using the macroscopic film properties.  Thus, when $L^i$ is determined by exchange coupling (probability $P_E$), we assign $L^i$ based on the average order parameter of the surrounding grains, weighted by border length: $L^i={\mathrm{sign}}\,[\sum_{j\in \mathcal{N}^i} L^j b^{i,j}]$, where $\mathcal{N}^i$ is the set of neighbours of grain $i$ and $b^{i,j}$.

To simulate the experiment on the mesh, we first cool with $\eta=0$ to a measurement temperature $T_{\mathrm{start}}$ and record $\{L^i_{\rm{init}}\}$.  Then, we heat to a temperature $T_{\rm{cycl}}$, resetting $L$ for all grains with $T_{\rm{crit}}<T_{\rm{cycl}}$.  Finally, we again cool to $T_{\mathrm{start}}$ and record $\{L^i_{\rm{final}}\}$.  We then calculate $B_{\rm{NV}}(x,y,h_{\rm{NV}})$ resulting from $\{L^i_{\rm{init}}\}$ and $\{L^i_{\rm{final}}\}$ (via Fourier propagation, as above), and correlate the initial and final magnetic field images, just as is done in the the experiment:
\begin{equation}\label{eq:CorrelationFunction}
C=\frac{\int B_{\rm{NV}}^{\rm{init}}(x,y)\,B_{\rm{NV}}^{\rm{final}}(x,y)\,dxdy}{\left(\left[\int (B_{\rm{NV}}^{\rm{init}}(x,y))^2\,dxdy\right]\left[\int (B_{\rm{NV}}^{\rm{final}}(x,y))^2\,dxdy\right]\right)^{1/2}}.
\end{equation}

This correlation function effectively minimizes the contribution of any parts of the film that are not ordered at $T_{\mathrm{start}}$.  Due to experimental constraints, for these experiments $T_{\rm{start}}=\SI{299.5}{\kelvin}$, which is within the range of $T_{\rm{crit}}$ for our film.  Therefore, a significant fraction of the film is not ordered at $T_{\mathrm{start}}$.  We take this into account in our simulations via the local $T_{\rm{crit}}^i$ values, by assigning a local magnetic moment density to each grain, $\sigma^i=\sigma_{\rm{sat}}(1-(T_{\rm{start}}/T_{\rm{crit}}^i))^{\beta}$, where $\sigma_{\rm{sat}}=15\,\mu_{B}/\rm{nm}^2$ and $\beta=0.35$.

At the end of the simulation we thus have the initial and final order parameter and the local moment density for each grain, from which we calculate the field in the plane of the NV centre, using the Fourier propagation method described above.  We then take the simulated fields before and after temperature cycling, add Gaussian random noise at the level of the field measurement noise in our experiment, and calculate $C$ as in Eq.\,\ref{eq:CorrelationFunction}.  The simulated value of $C(T_{\mathrm{cycl}})$ is plotted for $P_E=0.0, 0.33, 0.67, 1.0$ in Fig.\,4d.

\bibliographystyle{apsrev4-1}
\bibliography{BibCr2O3}

\end{document}